\documentclass[a4paper,oneside]{saip} 

\usepackage{comment}

\usepackage{parskip}

\begin{document} 

\title{From One to Eight: Supersymmetry Restoration in Lattice 3D ${\cal N} = 4$ Super Yang--Mills} 
\author{Dario van den Berg\,\orcidlink{0009-0003-0196-9796}$^{1}$ and
Anosh Joseph\,\orcidlink{0000-0003-4288-8207}$^{1}$}

\affil{$^{1}$\, National Institute for Theoretical and Computational Sciences, \\ School of Physics, and Mandelstam Institute for Theoretical Physics,\\ University of the Witwatersrand, Johannesburg, Wits 2050, South Africa}

\email{\url{dario.vandenberg@gmail.com}}

\begin{abstract}
Topological twisting provides a powerful framework for constructing lattice formulations of supersymmetric gauge theories. In three dimensions, a twisted version of ${\cal N} = 4$ super Yang--Mills theory can be discretized so that one nilpotent scalar supersymmetry is preserved exactly at nonzero lattice spacing. The remaining seven supersymmetries are broken by lattice artifacts of ${\cal O}(a)$, where $a$ is the lattice spacing. An important question is whether these supersymmetries are automatically restored in the continuum limit $a \to 0$, or whether fine-tuning of the lattice couplings is required. In this work, we derive the additional twisted supersymmetries by combining discrete $R$-symmetries of the continuum theory with the action of the scalar supercharge. This construction suggests that restoration of rotational symmetry in the continuum limit implies restoration of $R$-symmetry, leading to an automatic enhancement to the full ${\cal N} = 4$ supersymmetry without further tuning. These results may enable nonperturbative lattice studies of three-dimensional supersymmetric gauge theories relevant to string theory and mirror symmetry.
\end{abstract}

\section{Introduction}

Lattice field theory provides the only systematically improvable non-perturbative framework for studying strongly coupled quantum field theories. 
Extending these methods to supersymmetric gauge theories, however, is considerably more challenging than for ordinary gauge theories. 
A naive lattice discretization explicitly breaks the supersymmetry algebra since it contains the generators of infinitesimal spacetime translations, which are absent on a discrete lattice. 
Recovering the target continuum theory, therefore, generally requires the fine-tuning of multiple counterterms. 

A major advance in lattice supersymmetry came through the introduction of topologically twisted formulations of supersymmetric gauge theories. 
Following the twist, one of the supersymmetry generators becomes a nilpotent scalar supercharge $Q$, allowing an exact supersymmetry to be preserved even at nonzero lattice spacing. 
This exact lattice supersymmetry further constrains the structure of possible radiative corrections on the lattice. 
The approach of twisting forms the basis of modern lattice formulations of supersymmetric Yang--Mills theories \cite{Catterall:2009it, Joseph:2015xwa, Joseph:2023vja, Schaich:2022xgy}. 

Although the scalar supersymmetry survives discretization, the remaining twisted supercharges are explicitly broken by lattice artifacts. 
An important question is therefore whether these supersymmetries are automatically restored in the continuum limit or require additional fine-tuning. 
Earlier studies argued that restoration of an appropriate subset of discrete $R$-symmetries is sufficient to recover the full supersymmetric continuum theory \cite{Catterall:2013roa, Catterall:2014vka}. 
The geometric origin of these symmetries, however, has remained less transparent. 

In these proceedings, we revisit this question for the three-dimensional twisted formulation of $\mathcal N = 4$ super Yang--Mills theory. 
We show that the non-scalar twisted supercharges are naturally generated by conjugating the exact scalar supercharge with discrete automorphisms of the twisted theory. 
This construction leads to a simple geometric interpretation of lattice supersymmetry. 
The automorphisms interchange differential forms of different degrees and therefore cannot be realized as exact local gauge-covariant symmetries within the standard geometric
discretization. 
Consequently, the breaking of the non-scalar supersymmetries at finite lattice spacing is understood as a geometric obstruction rather than a dynamical effect. 

The above perspective also tells us the mechanism through which the full twisted supersymmetry algebra is expected to emerge in the continuum limit. 
Since the scalar supercharge is preserved exactly at finite lattice spacing, restoration of the discrete automorphism structure immediately reconstructs the remaining twisted supercharges through their conjugation relations. 
Supersymmetry restoration is therefore recast as a problem of geometric restoration, thereby providing a unified interpretation of the role of discrete $R$-symmetries in twisted-lattice supersymmetric gauge theories. 

\section{Twisted Lattice $\mathcal N = 4$ Super Yang--Mills Theory} 

The lattice formulation considered in this work is based on the topological twisting of three-dimensional $\mathcal N = 4$ super Yang--Mills theory. 
Instead of discretizing the supersymmetric theory directly, we can reorganize the fields and supersymmetry generators into new representations (of a twisted rotation group) that are compatible with a geometric lattice discretization. 
The resulting theory preserves one nilpotent scalar supersymmetry exactly at a nonzero lattice spacing. 
The twisted construction provides the foundation for modern lattice formulations of supersymmetric gauge theories \cite{Catterall:2009it, Catterall:2013roa}. 

The twisted theory considered here may be obtained by dimensional reduction of six-dimensional $\mathcal N = 1$ super Yang--Mills theory, followed by the Blau--Thompson twist \cite{Blau:1996bx}. 
Under this twist, the bosonic fields combine into a complexified gauge field $\mathcal A_a = A_a + i B_a$, where $A_a$ denotes the three-dimensional gauge field and $B_a$
originates from the scalar fields obtained through dimensional reduction. 
The fermionic degrees of freedom reorganize into antisymmetric tensor fields, 
\begin{equation} 
\left\{ \eta,\, \psi_a,\, \chi_{ab},\, \theta_{abc} \right\}, 
\end{equation} 
corresponding to differential forms of degree $0$, $1$, $2$, and $3$, respectively. 

The supersymmetry generators undergo an analogous decomposition into 
\begin{equation}
\left\{ Q, \, Q_a, \, Q_{ab}, \, Q_{abc} \right\}, 
\end{equation} 
where the scalar supercharge satisfies the nilpotency condition $Q^2 = 0$. 
The twisted supercharge $Q$ does not produce spacetime translations, unlike the conventional supersymmetry generators, and thus, can be preserved exactly after discretization. 

The action of the scalar supercharge on the twisted fields is 
\begin{equation}
\begin{aligned} 
Q \, \mathcal A_a &= \psi_a, & Q \, \psi_a &= 0, \\ 
Q \, \overline{\mathcal A}_a &= 0, & Q \, \chi_{ab} &= - \overline{\mathcal F}_{ab}, \\ 
Q \, \eta &= d, & Q \, d &= 0, \\ 
Q \, B_{abc} &= \theta_{abc}, & Q \, \theta_{abc} &= 0.
\end{aligned} 
\end{equation}
In the above, $d$ and $B_{abc}$ are auxiliary fields - they ensure off-shell closure of the supersymmetry algebra. 

Following the Dirac--K\"ahler construction, the twisted fermions are naturally associated with geometric objects of the lattice. 
Scalars reside on sites, one-forms on links, two-forms on plaquettes, and three-forms on elementary cubes. 
This geometric assignment preserves gauge invariance while eliminating the fermion doubling problem without introducing additional degrees of freedom \cite{Rabin:1981qj, Catterall:2013roa}. 

The lattice action consists of a $Q$-exact term together with a $Q$-closed topological term, ensuring exact invariance under the scalar supersymmetry at finite lattice spacing. 
In contrast, the remaining twisted supercharges are not manifest in the lattice construction. 
Understanding their origin and their role in the continuum limit is the main subject of this work. 

\section{Discrete Automorphisms and the Twisted Supercharges} 

The twisted formulation contains, in addition to the nilpotent scalar supercharge $Q$, a vector supercharge $Q_a$, an antisymmetric tensor supercharge $Q_{ab}$, and a three-form supercharge $Q_{abc}$. 
While the scalar supersymmetry is manifest in the lattice construction, the origin of the remaining supercharges is less transparent. 

A key observation is that these non-scalar generators are not independent. 
Instead, they are obtained from the scalar supercharge by conjugation with a set of discrete automorphisms of the twisted theory. 
Denoting these automorphisms by $R_a$, $R_{ab}$, and $R_{abc}$, the corresponding supersymmetry generators are given by
\begin{equation}
\label{eq:conjugation_relations}
Q_a = R_a Q R_a^{-1}, ~~ Q_{ab} = R_{ab} Q R_{ab}^{-1}, ~~ Q_{abc}  = R_{abc} Q R_{abc}^{-1}.
\end{equation} 

Since each $R$ is a symmetry of the continuum action, the conjugated operators are automatically symmetries as well. 
Furthermore, because they are related to the nilpotent scalar supercharge through similarity transformations, they inherit the same algebraic structure. 
The complete set of eight twisted supercharges is therefore generated from a single fundamental operator together with the discrete automorphism group. 

The action of the automorphisms is naturally understood in the language of differential forms. 
Rather than preserving the form degree, they interchange different sectors of the twisted complex. 
They map scalars into vectors, vectors into antisymmetric tensors, and higher-degree forms into one another. 
Consequently, the automorphisms not only relate different fermionic fields, but also connect the corresponding supersymmetry generators. 

This viewpoint provides a simple conceptual picture of the twisted supersymmetry algebra. 
Instead of looking at $Q$, $Q_a$, $Q_{ab}$, and $Q_{abc}$ as independent generators, we can treat them as different realizations of the same underlying supersymmetry; they are related through discrete transformations acting on the twisted differential-form complex. 
The automorphism structure, therefore, constitutes the organizing principle of the complete twisted supersymmetry algebra. 

This observation has an important consequence for lattice formulations. 
Since the scalar supercharge $Q$ is preserved exactly, the realization of the remaining supersymmetries is reduced to the question of whether the discrete automorphisms can also be realized on the lattice. 
As we discuss in the next section, the standard geometric discretization introduces a fundamental obstruction to this construction. 

\section{Geometric Obstruction on the Lattice} 

The standard lattice formulation of twisted supersymmetric gauge theories is based on a geometric discretization in which each field is assigned to a lattice cell according to its degree as a differential form. 
Scalar fields live on lattice sites, one-form fields on oriented links, two-forms on plaquettes, and three-forms on elementary cubes. 
This assignment has two advantages: it preserves gauge covariance and provides a natural realization of the Dirac--K\"ahler construction. 

The scalar supersymmetry $Q$ is compatible with this geometric structure. 
The $Q$-transformations map fields into other fields living on the same geometric object. 
As a result, the supersymmetry transformations commute with lattice gauge transformations, thus, allowing the scalar supercharge to remain an exact symmetry at nonzero lattice spacing. 

The situation changes qualitatively for the remaining twisted supercharges. 
Since they are generated through the automorphisms introduced in the previous section, they necessarily inherit the action of those automorphisms on the differential-form complex. 
They, therefore, map fields between different geometric sectors of the lattice. 

For example, the vector supercharges satisfy the schematic transformations 
\begin{equation} 
Q_a \, {\cal U}_b \sim \delta_{ab} \, \eta, \qquad Q_a \, \overline{\cal U}_b \sim \chi_{ab}, 
\end{equation} 
where ${\cal U}_b$ is a link field, $\eta$ is a site field, and $\chi_{ab}$ is a plaquette field. 
For the tensor supercharges $Q_{ab}$ and $Q_{abc}$ similar transformations occur. 
In every case, the supersymmetry transformation relates fields associated with different lattice cells. 

This observation immediately identifies the origin of the obstruction. 
Gauge covariance on the lattice depends explicitly on the geometric location of the fields. 
Fields living on sites, links, plaquettes, and three-cells obey different gauge transformation laws since they are associated with different parallel transport paths. 
A local symmetry that interchanges these geometric sectors, therefore, cannot commute with lattice gauge covariance. 

The absence of the non-scalar supersymmetries at finite lattice spacing is thus not a dynamical effect. 
Radiative corrections, strong-coupling effects, or anomalies do not produce it. 
Rather, it is a purely geometric consequence of combining three ingredients: 
\begin{enumerate} 
\item the organization of the twisted fields into differential forms, 
\item the assignment of these forms to lattice cells of different dimensions, and 
\item the implementation of gauge covariance through parallel transport. 
\end{enumerate} 

Any lattice formulation sharing these three features necessarily encounters the same obstruction. 

This result provides a geometric interpretation of the explicit breaking of the non-scalar supersymmetries. 
The twisted supersymmetry algebra itself remains perfectly well defined in the continuum theory. 
What fails on the lattice is the compatibility between the automorphisms of the differential-form complex and the geometric realization of gauge covariance. 
The obstruction, therefore, lies not in the supersymmetry algebra but in its realization within the standard geometric discretization. 

The existence of this obstruction naturally raises the question of how the full supersymmetry algebra is nevertheless recovered in the continuum limit. 
As we discuss in the next section, the answer lies in restoring the underlying automorphism structure as the lattice spacing is taken to zero. 

\section{Continuum Restoration of Supersymmetry} 

The geometric obstruction discussed in the previous section is a consequence of the finite lattice spacing. 
It originates from the unit-cell structure of the lattice, where sites, links, plaquettes, and three-cells are distinct geometric objects that carry different gauge-transformation properties. 
As the lattice spacing is taken to zero, this distinction gradually disappears, and the lattice approaches a smooth continuum manifold. 

In the continuum theory, the twisted fields are naturally interpreted as differential forms on a common geometric space. 
The discrete automorphisms, therefore, act within a unified differential-form complex, instead of acting between distinct lattice cells. 
The geometric obstruction present at finite lattice spacing is consequently expected to vanish in the continuum limit. 

The lattice construction preserves the scalar supercharge $Q$. 
If the discrete automorphisms $R_a$, $R_{ab}$, and $R_{abc}$ are restored as symmetries of the long-distance theory, then the remaining non-scalar supercharges follow immediately through the conjugation relations, Eq. \eqref{eq:conjugation_relations}. 
The complete twisted supersymmetry algebra is therefore recovered without requiring the non-scalar supercharges to be exact at finite lattice spacing. 

This restoration has a simple geometric interpretation. 
At finite lattice spacing, the automorphisms cannot be realized because they interchange fields that live in different geometric cells. 
In the continuum, however, the distinction between sites, links, plaquettes, and three-cells disappears, and the twisted fields once again form a single differential-form complex. 
The automorphism structure is then recovered together with the complete set of twisted supersymmetries. 

\begin{figure} 
\centering 
\includegraphics[width=0.5\linewidth]{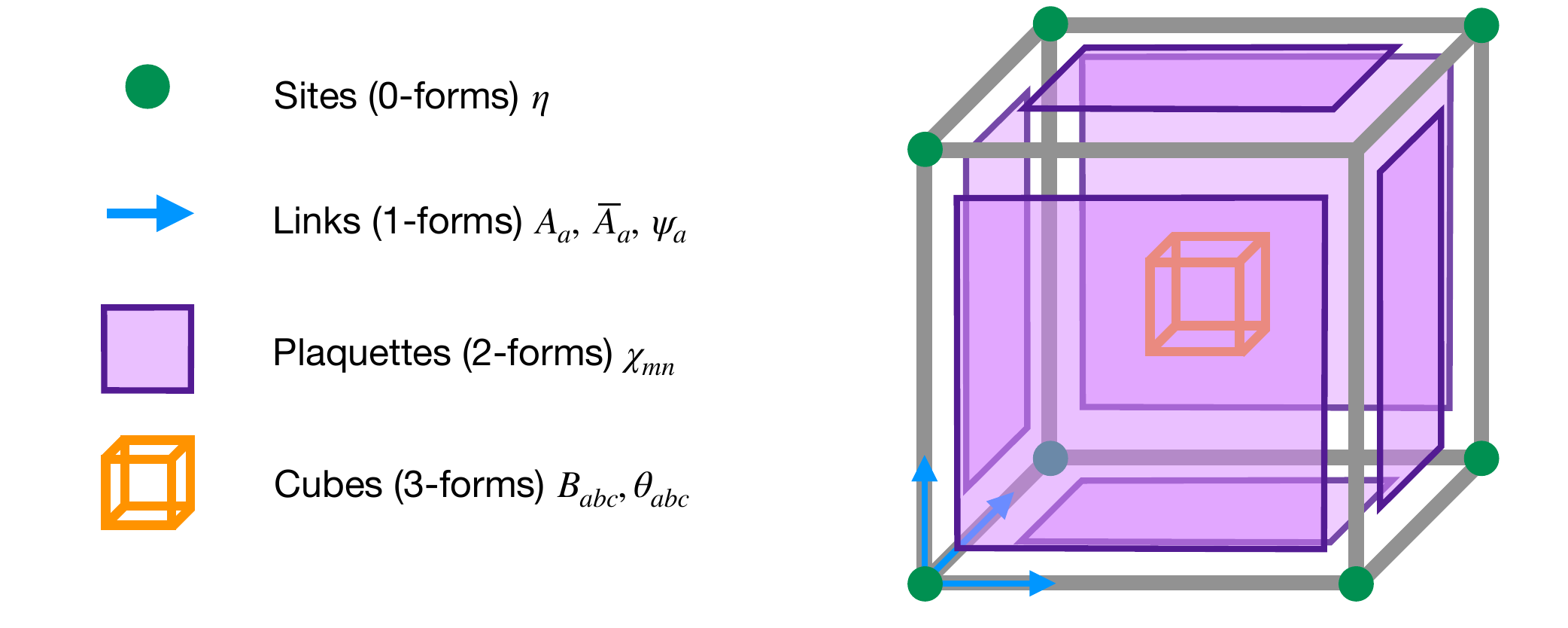} 
\caption{Geometric placement of the twisted fields on the lattice. 
The scalar supercharge preserves the geometric assignment of the fields, while the discrete automorphisms relate fields residing on different lattice cells. 
In the continuum limit, these geometric distinctions disappear, allowing the full twisted supersymmetry algebra to be recovered.} 
\label{fig:lattice} 
\end{figure}

The geometric organization of the twisted fields is illustrated in Fig.~\ref{fig:lattice}. 
The scalar supercharge acts within each geometric sector, whereas the discrete automorphisms relate fields of different form degree. 
Once these automorphisms are restored in the continuum, they generate the remaining twisted supercharges through the conjugation relations above. 

The restoration of the automorphism structure is equivalent to the restoration of the complete twisted supersymmetry algebra. 
In three dimensions, we can parametrize the complete set of supersymmetry transformations by a scalar $\delta \kappa_0$, a vector $\delta \kappa_a$, an antisymmetric tensor $\delta \kappa_{ab}$, and a three-form $\delta \kappa_{abc}$. 
We can write the corresponding supersymmetry variation in a compact form  
\begin{equation} 
\delta_\kappa = \delta \kappa_0 Q + \delta \kappa_a Q_a + \delta \kappa_{ab} Q_{ab} + \delta \kappa_{abc} Q_{abc}. 
\end{equation} 

This operator, when it acts on the twisted fields, reproduces the complete continuum supersymmetry transformations, 
\begin{eqnarray}
\delta_\kappa {\cal A}_m &=& \delta_{\kappa_0} \psi_m + \frac12 \delta_{\kappa_m} \eta + \delta_{\kappa_{ab}} \frac12 \theta_{abm} - \delta_{\kappa_{mbc}} \chi_{bc}, \\
\delta_\kappa \overline{\cal A}_m &=& \delta_{\kappa_{ab}} (\delta_{am} \psi_b - \delta_{bm} \psi_a), \\
\delta_\kappa \psi_m &=& \delta_{\kappa_a} \Big( \frac12 \delta_{am} d_a + (1 - \delta_{am}) [\overline{\cal D}_a, {\cal D}_m] \Big) + \delta_{\kappa_{abm}} \overline{\cal F}_{ab}, \\
\delta_\kappa \chi_{mn} &=& - \delta_{\kappa_0} \overline{\cal F}_{mn} + \delta_{\kappa_{ab}} \left(\frac12 (\delta_{am} \delta_{bn} - \delta_{bm} \delta_{an}) d_{mn} + \delta_{am} [\mathcal D_b, {\overline{\mathcal D}}_n] - \delta_{bm} [\mathcal D_a, {\overline{\mathcal D}}_n]\right), \\
\delta_\kappa \eta &=& \delta_{\kappa_0} d + 2 \delta_{\kappa_{ab}} {\cal F}_{ab}, \\
\delta_\kappa B_{abc} &=& \delta_{\kappa_0} \theta_{abc} - \delta_{\kappa_a} \chi_{bc} - 2 \delta_{\kappa_{ab}} \psi_c + \delta_{\kappa_{abc}} \eta, \\
\delta_\kappa \theta_{mbc} &=& - \delta_{\kappa_m} {\cal F}_{bc} + \delta_{\kappa_{mbc}} d_{mbc}.
\end{eqnarray} 

The auxiliary bosonic fields $(d, d_a, d_{ab}, d_{abc})$ allow us to close the supersymmetry algebra off shell. 
They transform trivially, $ \delta_\kappa d = \delta_\kappa d_a =
\delta_\kappa d_{ab} = \delta_\kappa d_{abc} = 0$. 

The transformations above represent the explicit realization of the full twisted supersymmetry algebra that is expected to emerge in the continuum limit. 
The recovery of non-scalar supercharges follows from the restoration of the discrete automorphism structure, instead of from requiring that each non-scalar supersymmetry be preserved independently at finite lattice spacing. 

This perspective also clarifies the role of the discrete $R$-symmetries that appear in earlier analyses of twisted lattice supersymmetry \cite{Catterall:2013roa, Catterall:2014vka}. 
Those works argued that the restoration of an appropriate subset of discrete $R$-symmetries is sufficient to recover the full supersymmetric continuum theory. 
From the present viewpoint, these discrete symmetries are nothing but the automorphisms that generate the non-scalar supercharges. 
Their restoration, therefore, reconstructs the complete twisted supersymmetry algebra through the conjugation relations discussed above. 

We emphasize that the restoration of the automorphism structure should be regarded as a physical expectation rather than a mathematical theorem. 
Demonstrating this restoration requires either a renormalization-group analysis or direct numerical evidence from lattice simulations. 
Nevertheless, the geometric framework presented here identifies the mechanism by which the complete supersymmetry algebra is expected to emerge in the continuum limit. 
It provides a natural interpretation of supersymmetry restoration in twisted lattice gauge theories. 

\section{Conclusions} 

In these proceedings, we have presented the problem of supersymmetry restoration in the lattice formulation of three-dimensional twisted $\mathcal N = 4$ super Yang--Mills theory from a geometric perspective. 
The main observation we arrived at is that the non-scalar twisted supercharges are not independent generators of the supersymmetry algebra; we can obtain them through a conjugation of the exact scalar supercharge with the discrete automorphisms of the twisted theory. 

This viewpoint yields a natural explanation for the absence of the non-scalar supersymmetries at finite lattice spacing. 
Within the standard geometric discretization, the automorphisms necessarily interchange fields living on different lattice cells. 
The gauge covariance is constrained to the geometric orientations of the fields, and thus, we cannot realize these automorphisms as exact local lattice symmetries. 
We can understand the resulting breaking of the non-scalar supersymmetries as a geometric obstruction instead of a dynamical effect. 

We see that the same geometric picture also suggests a simple mechanism for recovering the full supersymmetry algebra as we take the continuum limit of the theory. 
Since the scalar supercharge is preserved exactly at finite lattice spacing, the restoration of the automorphism structure reconstructs the remaining twisted supercharges through their conjugation relations. 
In this sense, restoration of the full twisted supersymmetry algebra may be viewed as a consequence of geometric restoration as the continuum limit is approached. 

The framework presented here provides a complementary perspective on the role of discrete $R$-symmetries in twisted lattice supersymmetry. 
It suggests that the emergence of an automorphism structure should be a useful diagnostic for the restoration of continuum supersymmetry. 
Establishing this mechanism quantitatively through numerical lattice simulations and renormalization-group analyses \cite{Catterall:2011pd} remains an interesting direction for future work. 

More broadly, twisted lattice formulations provide one of the few systematic nonperturbative approaches to supersymmetric gauge theories. 
A better understanding of the mechanism by which supersymmetry is restored in the continuum, therefore, strengthens the theoretical foundations of future numerical studies of three-dimensional $\mathcal N = 4$ super Yang--Mills theory. 
We hope that such studies provide valuable insights into strongly coupled supersymmetric field theories and their connections to string theory, mirror symmetry, and aspects of gauge/gravity duality. 

\section{Aknowledgements}

DVDB would like to thank the SA-CERN Excellence Programme, through NRF iThemba LABS, for the financial support that made the presentation of this work possible. 
AJ was supported in part by the Start-up Research Grant from the University of the Witwatersrand, South Africa. 
The authors also gratefully acknowledge support from the National Institute for Theoretical and Computational Sciences (NITheCS), the Mandelstam Institute for Theoretical Physics (MITP), and the School of Physics at the University of the Witwatersrand.

\bibliographystyle{IEEEtran}
\bibliography{biblio}

\end{document}